\documentclass[twocolumn,prl,showpacs]{revtex4}
\usepackage{graphicx,epsfig}

\def\be{\begin{equation}}
\def\ee{\end{equation}}

\begin{document}

\title{Polarization of interacting bosons with spin}

\author{Eli Eisenberg$^{1,2}$ and Elliott H. Lieb$^{1,}$}
\affiliation{$^1$Department of Physics, Princeton
University, P.O.B. 708, Princeton, NJ 08544, USA} 
\affiliation{$^2$NEC Research
Institute, 4 Independence Way, Princeton, NJ 08540,
USA}

\date{July 1, 2002}

\begin{abstract}
We prove that in the absence of explicit spin-dependent forces one of
the ground states of interacting bosons with spin is always fully
polarized.  Generally, this state is degenerate
with other states, but one can specify the exact degeneracy. For $T>0$
the magnetization and zero-field susceptibility exceed that of a pure
paramagnet.  The results are relevant to experimental work on triplet
superconductivity and condensation of atoms with spin. They eliminate
the possibility, raised in some theoretical speculations, that the
ground state or positive temperature state might be antiferromagnetic.

\end{abstract}


\maketitle

Bosons with spin first appeared in condensed matter physics in the
theory of $^3$He superfluidity, which is due to spin-triplet $p$-wave
pairing of the Helium atoms \cite{he}. More recently they were
discussed in connection with the rising interest in triplet
superconductivity \cite{ts1,ts2,ts3,ts4} and multicomponent
Bose-Einstein condensation \cite{bec,bec2,bec3}. The renewed
fascination with spin-triplet superconductivity was generated by
recent experimental evidence for triplet pairing in heavy fermion
systems \cite{ts1}, organic conductors \cite{ts2}, and the layered oxide
${\rm Sr_2RuO_4}$ \cite{ts3,ts4}. Recent advances in the study of the
Bose-Einstein condensation phenomenon enabled condensation of bosons
with non-zero spin (spin-1 in particular), such as $^{23}$Na and
$^{87}$Rb atoms, by confining them in optical traps rather than the
usual magnetic traps, thus preserving their spinor nature \cite{bec,bec2}.

Moreover, it is possible to have condensates of bosons with 
other internal degrees of freedom which behave like spin, thus realizing
``spin-1/2'' bosons. For example, a two component Bose gas was produced 
in a magnetically trapped $^{87}$Rb condensate
by rotating two hyperfine states into 
each other, creating an $SU(2)$ symmetry \cite{bec3}.

All these experimental achievements have generated new interest in the
study of interacting spinor bosons. In particular, some tantalizing
hints for a possible connection between the onset of triplet
superconductivity and the appearance of internal magnetic
moments \cite{ts3} call for better understanding of the magnetic
properties of strongly interacting bosons.  We report here an exact
and general result, stating that for a generic Hamiltonian of bosons
with spin, a fully polarized state is among the ground states -- as long
as there are no explicit spin-dependent forces -- however complicated the
many-body interaction potential might be.  
We can also classify all other ground states. 
The conclusion is that the ground state susceptibility is infinite.
We further show that the positive temperature magnetization and 
zero-field susceptibility
exceed those of non-interacting distinguishable (Maxwell-Boltzmann) spins.

Recent works \cite{2body} analyzed the two body interaction for
spin-$1$ atoms in an optical trap. Using an effective low energy approach,
the interaction term was parameterized there by
\begin{equation}
\label{2b}
V({\bf r}_i-{\bf r}_j)=\delta({\bf r}_i-{\bf r}_j)(c_0+c_2{\bf S}_1\cdot
{\bf S}_2)
\end{equation}
where the coefficients $c_0$ and $c_2$ are related to the $s$-wave
scattering lengths in the singlet and triplet channels. The
possibility of an antiferromagnetic coupling $c_2>0$ has attracted
much attention in the past few years.  Experimental estimates for these
scattering lengths suggest that $c_2>0$ for a condensate of $^{23}$Na
atoms \cite{nascat}. In such a case, the energy is minimized by a
polar state with $\langle{\bf S}\rangle=0$. We point out that in view
of the following exact results this scenario can be ruled out as long
as the full many-body interaction is spin-independent.  In fact, as we
explain later, the theorem below shows that the effective two-body
spin-dependent interaction described in Eq. (\ref{2b}) can not
describe the low-energy behavior of bosons with spin-independent
interaction, {\it no matter what the sign of $c_2$ is}.  
Therefore, a bosonic spin-spin term of the type (\ref{2b}) can only be 
justified by going beyond a spin-independent bosonic Hamiltonian, 
either by introducing explicit spin-dependent forces, or by taking into
account the underlying fermionic physics of the atoms.  
This situation should be contrasted with the
fermionic case in which effective exchange interactions, ferromagnetic
or antiferromagnetic, do come about from spin-independent
Hamiltonians.

The Hamiltonian of a Bose gas of $N$ bosons, and without
spin-dependent forces, is given by
\begin{equation}  \label{ham}
H=-\sum_{i=1}^N \frac{1}{2m}\nabla^2_i + v({\bf x}_1,{\bf x}_2,\cdots,
{\bf x}_N)
\end{equation}
where $m$ is the mass of the bosons, and ${\bf x}_1,{\bf x}_2,\cdots,
{\bf x}_N$ are the spatial coordinates of the bosons. $v$ is a
spin-independent, totally symmetric, potential, which includes any
confinement, disorder, and $k$-body interactions.  We look for the
lowest energy state that is totally symmetric under permutations of
the space-spin indices. To find it, we temporarily ignore both the
spin of the particles and the identity of the bosons at first, and
look for the ground state of $N$ {\it non-identical particles}, i.e.,
particles without spin and without any symmetry restriction on the
wave-function. We seek the absolutely lowest state of $H$, $\psi({\bf
x}_1,{\bf x}_2,\cdots, {\bf x}_N)$.

It is known (e.g., \cite{Lieb63}) that this ground state is unique 
(up to an overall phase) and  satisfies
\begin{equation}
\label{positive}
\psi({\bf x}_1,\cdots,{\bf x}_N) 
> 0\quad\quad{\rm for\ all\ \ }
{\bf x}_1,\cdots,
{\bf x}_N
\end{equation}
(unless $v$ is so repulsive that it prevents particles from changing
places by a continuous path -- which would be non-physical). This
result holds for all boundary conditions (Dirichlet, von-Neumann,
periodic).  Eq. (\ref{positive}) follows immediately from the fact
that $\psi$ can be chosen to be real (because $\psi^*$ is also a
ground state and hence $\psi \pm \psi^*$ are ground states), and from
the fact that the variational energy of the function $|\psi|$ then
equals that of $\psi$ itself \cite[Eq. (6.17.2)]{LL}. Hence $|\psi|$
must be a ground state.  From this we conclude that if $\psi$ is a
real ground state, either $\psi=\alpha|\psi|$, and thus
(\ref{positive}) is satisfied, or $|\psi|-\psi$ is a ground state
that is zero on a set of positive measure, which is impossible unless
$v$ has the pathology mentioned above. If, now, $\phi$ is another real
ground state that is orthogonal to $\psi$ then $\phi = |\phi|$ up to a
phase (as we just proved). But two positive functions cannot be
orthogonal, so $\psi $ is unique.

Equation (\ref{positive}) implies that $\psi$ must be totally
symmetric. I.e., if we sum $\psi$ over all permutations we obtain a
function that is (a) symmetric, (b) non-zero, and (c) a ground
state. By uniqueness, this symmetric function must equal $\psi$ up to
a constant.

We now return to our identical spin-full boson problem, and define
$$ 
\phi({\bf x}_1,\cdots,{\bf x}_N;\sigma_1,\cdots,\sigma_N) \equiv
\psi({\bf x}_1,\cdots,{\bf x}_N)|\uparrow\uparrow\uparrow\cdots\rangle
$$
This function is totally symmetric in both spin
and spatial indices, and thus is a valid wave-function for bosons
(with spin).  As we have shown it minimizes the energy regardless of
symmetry restriction, and is, therefore, a ground state of the boson
system.

This ground state has spin angular momentum $J=\nolinebreak NS$, where $S$ is each 
boson's spin, and carries the usual $(2SN+1)$-fold degeneracy. 
There will be other ground states, but each must be a spin function times
the unique spatial function we have just described.
The question of the ground state degeneracy thus reduces to the 
following question:
How many different spin functions are there that are totally symmetric
(up to the trivial $2J+1$-fold degeneracy)?
The answer is as follows. 

\noindent
(i). If $S=1/2$ (``spin-1/2'' bosons \cite{note}) 
then there is just one function, $J=N/2$, and
the ground state is therefore non-degenerate (except for the trivial
degeneracy). \\
\noindent
(ii). If $S=1$ there are functions with $J=N, \, N-\nolinebreak 2,\linebreak 
N-4, ...$
and each of these appears exactly {\it once}. Wave-functions 
with $J=N-1,\,N-3,\,N-5, ...$ do not appear in the ground state.
This is  contrary to what one would 
have for a paramagnet in which all $J$ values are degenerate, 
and $J$-values smaller than $N$ occur multiple times.\\
\noindent
(iii). Other values of $S$ can be similarly analyzed by studying 
representations of $SU(2S+1)$ or by studying $(2S+1)$-rowed Young's tableaux.

If a magnetic field term $H_M= -\mu h\sum_{i=1}^N \sigma^z_i$ 
(where $h$ is the field, and $\mu\sigma^z_i$ is the atomic magnetic moment
in the $\hat z$ direction, including the g-factor), 
is added to $H$ then the ground state 
energy shifts by $-\mu hNS$, i.e., the zero-field
susceptibility is infinite.

We now discuss the positive temperature case  and compare it with
pure paramagnetism. In the following, a pure
paramagnet means a set of $N$ non-interacting, distinguishable, spin-$S$
particles, which has no degrees of freedom besides the spin.
The pure paramagnet magnetization is, therefore,
\begin{eqnarray}\label{thm}
M^{\rm para}(T,h)\!\!\!\!&=&\!\! N\mu \, \frac{\sum_{\sigma = -S}^S
\sigma \exp[\beta \mu h \sigma]}
{\sum_{\sigma = -S}^S \exp[\beta \mu h \sigma]}\ \\
&=&\!\! \frac{N\mu}{2}\  \biggl[(2S+1)\coth\left(\frac{2S+1}{2}\beta\mu h
\right) \nonumber  \\ 
&& \qquad\qquad
- \coth\left(\frac{1}{2}\beta\mu h\right)\biggr], \nonumber
\end{eqnarray}
where  $\beta =1/k_BT$.

Our theorem about enhanced magnetization in the ground state can be
generalized to positive temperature $T$, as follows.

\bigskip
THEOREM: {\it For each $T$ and magnetic field $h>0$, 
the magnetization $M(T,h)$ is greater than the pure paramagnetic value 
$M^{\rm para}(T,h)$. Moreover, the zero-field susceptibility 
 $\partial M(T,h)/\partial h|_{h=0}$, also exceeds
that of a pure paramagnet, \, $\partial M^{\rm para}(T,h)/
\partial h|_{h=0}$.   } 

\bigskip
{\it Proof.}---
Let us denote a particle configuration $x_1,x_2, ...,x_N$ by $X$ and a
spin configuration (in the $\sigma^z$ basis) $\sigma_1, \sigma_2,
...\sigma_N$ by $\Sigma$.  Since spin does not enter $H$ except
through $H_M$ the (operator) Boltzmann factor has the kernel
$$
K(X,X';T)\exp\left[\, \beta \mu h S(\Sigma)\, \right]
\delta_{\Sigma,\Sigma'}
$$ 
where $S(\Sigma)=\sum_{i=1}^N\sigma_i^z$, 
$\delta_{\Sigma,\Sigma'}$ is the `Kronecker delta' 
$\Pi^N_{i=1}\delta_{\sigma_i,\sigma'_i}$
and $K$ is the (unsymmetrized) kernel of $\exp[\, -\beta H\, ]$.
The symmetrized Boltzmann factor is then
\be \label{BF}
{\mathcal{S}}_{X,\Sigma}\, K(X,X';T)
\exp\left[\,\beta \mu h S(\Sigma)
\, \right]  \ \delta_{\Sigma,\Sigma'}\ .
\ee
and ${\mathcal{S}}_{X,\Sigma}$ is the symmetrizer on 
the $X, \Sigma$ variables. 
(It is not necessary to symmetrize on the $X',\Sigma'$ since this will 
be automatic.)

The important point to notice is the $T>0$ analogue of
(\ref{positive}), namely, $K(X,X';T) >0$ for all $X,X'$ (outside of a
`hard-core' region). This can easily be seen by the path-space
(Wiener) integral for $\exp[\, -\beta H\, ](X,X')$ via the Feynman-Kac
formula -- or else via the Trotter product formula \cite{simon}. 

The Trotter formula states that $\exp[\, -\beta H\, ](X,X')$ is the limit
as $N\to \infty$ of
\begin{eqnarray}
\int A(X,X_1)B(X_1,X_2)A(X_2,X_3)B(X_3,X_4)\cdots &&\nonumber\\
\quad A(X_{2N-2},X_{2N-1})
B(X_{2N-1},X') dX_1\cdots dX_{2N-1} &&\label{trotter}
\end{eqnarray}
where
$A(X,Y)=\exp[\, -\beta H_0/N \, ](X,Y)$ (with $H_0=$ kinetic energy operator)
and $B(X,Y)=\exp[\, -\beta v/N \, ](X,Y) = \exp[\, -\beta v(X)/N \, ]
\delta (X-Y)$. The kernel $A$ is a Gaussian, $A(X,Y) = C_1 \exp[\, C_2
(X-Y)^2 \, ]$. Since all kernels are nonnegative, we see that the
multiple integral (\ref{trotter}) yields a nonnegative function. The
limit of nonnegative functions is nonnegative and, with some work one
can show that $K(X,X';T) >0$ for all $X,X'$ (outside of a `hard-core'
region).  In case $\exp[\,- v(X) \, ]$ is not integrable we must first
approximate it by a bounded function and then remove the
approximations at the end.  The positivity of the kernel is conserved
under symmetrization.

In fact, the only thing we really need is the weaker assertion that
$K_\pi (X;T)$, defined below, is nonnegative and is 
positive on a set of positive $X$ measure.

To calculate the partition function 
$Z(T,h)$ we set $X=X'$ and $\Sigma=\Sigma'$ (after applying
 ${\mathcal{S}}_{X,\Sigma}$) and then integrate over 
$X$ and sum over $\Sigma$
(i.e., $\sum_\Sigma=\Pi_{i=1}^N\sum_{\sigma_1 = -S}^S$).
We obtain an expression of the form
\begin{eqnarray} 
Z(T,h) &=&   \frac{1}{N!}\sum_\pi \int dX  K_\pi(X;T)
\nonumber \\    
&&\times \sum_\Sigma\ \exp\left[\beta \mu h S(\Sigma)\right]
\delta_{\pi(\Sigma),\Sigma}\ , \nonumber
\end{eqnarray}
where the first summation is over all permutations $\pi$ and 
$K_\pi (X;T)$ is a $\pi$-dependent function of $X$ and $T$ defined by
$K_\pi (X;T) = K(\pi X,\,  X;T)$.

To compute $M(T,h)$ let us
define
\begin{equation}\label{part}
Z_\pi(T,h) =  \sum_\Sigma
\exp\left[\beta \mu h S(\Sigma)\right]
\delta_{\pi(\Sigma), \Sigma} \  ,
\end{equation}
\begin{equation}\label{mpi}
M_\pi(T,h) =  \frac{\sum_\Sigma S(\Sigma)
\exp\left[\beta \mu h S(\Sigma)\right]
\delta_{\pi(\Sigma), \Sigma}} {Z_\pi(T,h)}  \ ,
\end{equation}
\begin{equation}
C_\pi(T,h) = Z_\pi(T,h)\int  K_\pi(X;T) dX \ .
\end{equation}
Then we can write
\begin{equation}\label{em}
M(T,h)= \frac{ \sum_\pi C_\pi(T,h) M_\pi(T,h) }
 {\sum_\pi C_\pi(T,h) } \ .
\end{equation}
Since $K(X,X';T)\geq 0$ and $>0$ on a set of positive $X$ measure,
$C_\pi (T,h)>0$, and we have that $M(T,h) > \min_\pi \
M_\pi(T,h)$. (The strict inequality $>$ follows from the facts that
$C_\pi (T,h)>0$ and that the numbers $M_\pi(T,h)$ are not all equal.)

Our problem reduces to deciding which $\pi$ will make (\ref{mpi}) as
small as possible.  It is the identity permutation, and this gives
$M^{\rm para}(T,h)$ as the lower bound. To see this we note that
$M_\pi(T,h)$ is the magnetization of distinguishable particles that
interact only through the constraint given by
$\delta_{\pi(\Sigma),\Sigma}$.  If $\pi$ is the identity, there is no
constraint, and one gets the pure paramagnetic magnetization
(\ref{thm}). For any other permutation $\pi$, $\delta_{\pi(\Sigma),
\Sigma}$ connects a group (or groups) of, say, $k>1$ spins. There is 
one group of size $k$ for each cycle of length $k$ in $\pi$.
This grouping 
only raises the magnetization by reducing the entropy of the low
moment states. In other words, the assertion follows from the
inequality 
$$  k\frac{\sum_{\sigma = -S}^S \sigma
\exp[\beta \mu h \sigma]}
{\sum_{\sigma = -S}^S \exp[\beta \mu h \sigma]}
<
\frac{\sum_{\sigma = -S}^S (k\sigma)
\exp[\beta \mu h (k\sigma)]}
{\sum_{\sigma = -S}^S \exp[ \beta \mu h (k \sigma)]}
$$ 
whose validity can be immediately seen by rewriting it as
$kM(T,h)<kM(T/k,h)$.
This inequality means that the magnetization of each group of 
$k$ spins is higher than that of $k$ independent spins. 
Since the magnetization of different groups is additive, we conclude
that the  constraint of tying together a group of spins
increases the magnetization; therefore, the minimal
$M_\pi(T,h)$ occurs at $\pi$ being the identity.

Finally, the fact that the $h=0$ susceptibility exceeds the
paramagnetic value follows from the two facts that $M(T,0)=0$ and
$M(T,h)$ is greater than the paramagnetic value for all $h$. Q.E.D.

\bigskip
{\it Some Generalizations.}--- 

(A) The previous results hold true even if there are several species
of bosons in the system, with possibly different masses, as long as
there are no fermions present. The only difference is that the
degeneracy of the ground state is the product of the degeneracies
found in (i), (ii), (iii) above for each species.

(B) The results hold for a `relativistic' boson system obtained by
replacing $p^2/2m$ by $\sqrt{p^2 +m^2} = \sqrt{-\nabla^2 +m^2}$. The
underlying reason is that the kernel of $\exp\left[-\beta\sqrt{p^2
+m^2}\right]$ is also positive \cite[Eq. (7.11.11)]{LL}. Lest the reader
think that positivity is a triviality, we remark that it does not hold
for $\exp\left[-\beta p^4\right]$.

(C) Our results hold in any dimension, of course. They also hold in
lattice models (e.g., sometimes called `Bose-Hubbard' models) in which
the $\Delta$ in the kinetic energy, $-\Delta = -\nabla^2 $, is replaced
by the discrete lattice Laplacian (= second difference operator) or
$\sum_{\langle i,j \rangle} a^\dagger_ia_j$ in the usual second quantized
notation. No regularity of the lattice is needed; indeed, the
`lattice' can be any general connected graph.

Several recent works have discussed boson models with an
antiferromagnetic coupling in the effective low-energy Hamiltonian,
resulting in a polar ground state.  Our theorem eliminates the
possibility of the Hamiltonian (\ref{ham}) producing an
antiferromagnetic state for bosons. Furthermore, it follows that the
low energy parameterization (\ref{2b}) of the Hamiltonian (\ref{ham})
{\it must have} $c_2=0$. To be specific, we discuss the spin-1 case,
but the situation is similar for bosons with higher spin.
The potential (\ref{2b}) with $c_2=0$ satisfies the conditions
of our theorem, and therefore its ground state 
for spin-1 bosons is degenerate, having states with 
$J=N, \, N-2,\, N-4,...$. 
Switching on the spin-spin interaction, with non-zero $c_2$,
would choose between these degenerate states, break the
degeneracy, and contradict the theorem. Therefore, a non-zero $c_2$ is
inconsistent with a correct low energy theory of the Hamiltonian
(\ref{ham}). Possible sources for such spin-spin interaction are then
the (rather weak) direct dipole-dipole magnetic interaction, or exchange 
terms coming from the electron transfer between the atoms, thus going
beyond the bosonic description of the system.

In a recent experiment \cite{bec2} the magnetic structure of a
Bose-Einstein condensate of spin-$1$ $^{87}$Rb atoms was explored. 
The authors studied the distribution of the different magnetic number 
$m$ values, but did not
look at the global polarization properties. Based on the above, we
suggest looking at the magnetization at finite field strength.
According to the paramagnetic lower bound (\ref{thm}), 
the magnetization at the experimental temperature should exceed half of
the maximal magnetic moment at fields as low as $1$G. One
can also rotate the macroscopic magnetic moment of the condensate to a
specific direction (by an appropriate  magnetic field), and then
(after switching off the field and allowing for relaxation) measure
the distribution of the $m$ values, quantized along the same axis. A
macroscopic moment should show up as a large fraction of the bosons in
the $m=1$ state. As discussed above, violation of the above predictions 
would indicate the importance of additional interaction terms, 
going beyond the Hamiltonian (\ref{ham}).

In conclusion, we have shown that fully polarized states are among the
ground states of interacting bosons.
Moreover, the finite temperature magnetization and zero-field
susceptibility were shown to be bigger than that of
Maxwell-Boltzmann (distinguishable) free spins.
Our results hold in the
presence of random on-site disorder and density-density interactions.
They hold even if there is no interaction potential at all, because
our theorems apply to the Hamiltonian $H=\sum p^2/2m$ as well.  They might be
relevant for recent experiments suggesting the formation of an
internal magnetic moment at the onset of spin-triplet
superconductivity, and to recent magnetic measurements of
Bose-Einstein condensates.  In particular, our ground state result
constrains effective low-energy theories for interacting bosons,
namely, it excludes the possibility of a spin-spin two-body term in
the absence of explicit spin-dependent interactions in the underlying
Hamiltonian.

\begin{acknowledgments}
We thank Karsten Held, Robert Seiringer, and Jakob Yngvason for many
useful comments. EHL thanks the National Science Foundation, grant PHY
0139984, for partial support of this work.
\end{acknowledgments}

{\it Note added in proof.--} After this paper was submitted for
publication we learned that similar mathematical results were obtained
by Andr\'as S\"ut\H o \cite{Suto}  in a study of ``cycle percolation''
in Bose gases. Our theorem that the pure paramagnetic
value is a lower bound to the magnetization for nonzero field 
was not given there, however.

\end{document}